\documentclass[prb,showpacs,twocolumn,superbib,floats,,superscriptaddress]{revtex4}
\usepackage{graphicx}
\usepackage{epsfig}
\usepackage{amsfonts}
\usepackage{amsmath}
\usepackage{natbib}
\usepackage{multirow}
\usepackage{bm}
\usepackage{epstopdf}
\usepackage{color}
\usepackage{ulem}

\begin{document}

\title{Effect of spin-orbit interaction on the excitonic effects 
in single-layer, double-layer, and bulk MoS$_2$}

\author{Alejandro Molina-S\'{a}nchez}
\affiliation{Physics and Materials Science Research Unit, University of Luxembourg, L-1511 Luxembourg, Luxembourg}
\affiliation{Institute for Electronics, Microelectronics, and Nanotechnology (IEMN), CNRS UMR 8520, Dept. ISEN, 59652 Villeneuve d'Ascq Cedex, France}

\author{Davide Sangalli}
\affiliation{MDM Lab, IMM, Consiglio Nazionale delle Ricerche, Via C. Olivetti, 2 I-20864 Agrate Brianza, Italy}

\author{Kerstin Hummer}
\affiliation{Faculty of Physics, Center for Computational Materials Science, University of Vienna, Sensengasse 8, A-1090 Wien, Austria}

\author{Andrea Marini}
\affiliation{Istituto di Struttura della Materia of the National Research Council, Via Salaria Km 29.3, I-00016 Monterotondo Stazione, Italy}

\author{Ludger Wirtz}
\affiliation{Physics and Materials Science Research Unit, University of Luxembourg, L-1511 Luxembourg, Luxembourg}
\affiliation{Institute for Electronics, Microelectronics, and Nanotechnology (IEMN), CNRS UMR 8520, Dept. ISEN, 59652 Villeneuve d'Ascq Cedex, France}

\date{\today}

\begin{abstract}

We present converged ab-initio calculations of the optical absorption spectra 
of single-layer, bi-layer, and bulk MoS$_2$. 
Both the quasiparticle-energy calculations (on the level of the GW approximation
) and the calculation of the absorption spectra (on the level of the 
Bethe-Salpeter equation) explicitly include spin-orbit coupling, using the 
full spinorial Kohn-Sham wave-functions as input. Without excitonic effects, 
the absorption 
spectra would have the form of a step-function, corresponding to the joint-density of states of a parabolic band-dispersion in 2D. This profile is deformed by a pronounced bound excitonic peak below the continuum onset. 
The peak is split by spin-orbit interaction in the case of single-layer and (mostly) by inter-layer interaction in the case of double-layer and bulk MoS$_2$. 
The resulting absorption spectra are thus very similar in the three cases but 
the interpretation of the spectra is different. Differences in the spectra
can be seen around 3 eV where the spectra
of single and double-layer are dominated by a strongly bound exciton.

\end{abstract}


\maketitle

\section{Introduction} 

The promising and interesting physical properties of graphene\cite{novoselov} have recently stimulated
active research in other atomically thin materials, alternative and/or complementary to graphene.\cite{coleman} 
Molybdenum disulfide (MoS$_2$), fabricated in its single-layer by means of 
mechanical exfoliation, exhibits a direct bandgap of 1.8-1.9 eV, contrary to the indirect band gap of its bulk
counterpart.\cite{splendiani2010,mak2010} Moreover, single-layer MoS$_2$ has also shown a mobility of 200 cm$^2/$Vs which
makes it appealing for the design of a new generation of more efficient transistors.\cite{kis2011} In the field
of spintronics, the absence of inversion symmetry in the crystal structure of
single-layer MoS$_2$ allows valley polarization by optical pumping with circularly polarized light.\cite{yao2008,xiao2012} This makes possible the design 
of devices based on  spin and valley control.\cite{zeng2012,mak2012} 
More recently, a remarkable Seebeck coefficient
has been measured in single-layer MoS$_2$, opening a new field of application for those materials.\cite{castellanos2013} Concerning the 
optical properties, the photoluminescence in single-layers has shown higher 
efficiency than in multi-layers or bulk which is attributed to the direct/
indirect band-gaps, respectively. The absorption spectra however 
are very similar in all the cases,\cite{splendiani2010,mak2010,castellanos2011} an issue not yet explained. The observed double-peak structure in the absorption
spectra can be connected to the splitting of the valence band maximum
around the high-symmetry point $K$. For single-layer MoS$_2$, this splitting 
was explained as a consequence of spin-orbit coupling which is a result
of the missing inversion symmetry.\cite{zhu11} For bilayer and bulk, this
splitting is mainly due to inter-layer interaction.\cite{lambrecht2012}

In this context, reliable \textit{ab initio} calculations of the absorption
spectra are necessary for providing the correct interpretation of the 
reported experimental results. However, the inherent complexity of the 
Bethe-Salpeter equation\cite{onida2002} that is usually used to describe 
the excitonic effects in the optical spectra seems to be the reason for
recent inadequately converged calculations. In Ref.~\onlinecite{rama} the 
excitonic binding energy is strongly overestimated due to a low k-point 
sampling and in Refs.~\onlinecite{singh} and~\onlinecite{feng12} spin-orbit interaction
was entirely neglected and the splitting of the excitonic peak is merely
due to an unconverged k-point sampling. 
The aim of our work is to provide well-converged optical spectra, 
in the framework of the Bethe-Salpeter equation {\it including the effects
of spin-orbit coupling}. This gives a reliable basis for the interpretation
of previous experimental works on single-, double-layer and bulk MoS$_2$. 
We show that the the optical spectrum corresponds essentially to a 
step-function that is the result of the joint-density of states for 
parabolic dispersion in two-dimensional systems. Excitonic effects shift
part of the oscillator strength into a discrete excitonic peak below
the continuum onset. The splitting of this excitonic peak can be
directly related to the splitting of the valence band maximum around
$K$ and is thus entirely due to spin-orbit coupling in the case of
the single-layer and mostly (but not entirely) due to inter-layer interaction
for double, and multi-layer MoS$_2$.

\section{Calculation Methods} 
Starting point of the calculation of optical spectra are the Kohn-Sham wave functions and energies calculated with density-functional theory (DFT) 
in the local-density approximation (LDA). We use the code {\tt ABINIT}\cite{gonze2002}
where wave-functions are expanded in plane-waves and core electrons are
simulated by norm-conserving relativistic pseudopotentials.\cite{hgh}
The plane-wave energy cutoff is 30 a.u. 
For Molybdenum, the 4s and 4p semi-core electrons are explicitly calculated
(in addition to the 4d and 5s valence electrons). 
This turns out to be crucial
for the proper calculation of the exchange-contribution to the self-energy
term in the GW calculations. As noted earlier,\cite{zhu11,lambrecht2012}
spin-orbit interaction is important for MoS$_2$ and we calculate
the spinor-wavefunctions as input for the following calculations on the 
level of many-body perturbation theory.

The inherent underestimation of the bandgap given by DFT is 
corrected with the GW method.\cite{hedin1970,onida2002}
We use the non-self consistent version (denoted as G$_0$W$_0$) without
updating the dielectric function in the screened Coulomb potential (W) or the 
wave-functions and energies in the Green's function (G). These calculations
are done with the {\tt Yambo} code.\cite{yambo} The dielectric function
$\epsilon_{G,G'}(\omega,q)$ is calculated using the plasmon-pole approximation\cite{ppa_note}.
50 G-vectors are used (for a vacuum distance of 40 a.u. between the
periodic images of the single-layer/double-layer calculations).
200 unoccupied bands are used in the integration of the self-energy term,
yielding converged band-gap corrections for single-layer, double-layer,
and bulk. The ${\bf k}$-point sampling is $18 \times 18 \times 1$ for 
single and double-layer and $18 \times 18 \times 3$ for bulk.
The value of the GW-correction to the band-gap depends on the
inter-layer distance in the periodic supercells approach. It increases
with increasing distance and converges
roughly as $1/d$ (where $d$ is the supercell dimension perpendicular to the
layer). This was shown for single layers of hexagonal boron nitride (hBN)\cite{wirtz2006} and for single-layers of MoS$_2$.\cite{komsa} At the same time,
the excitonic binding energy was also shown to increase roughly as $1/d$
such that the two effects cancel and the
resulting optical spectra hardly depend on the inter-layer distance.
In our calculations, we use $d=40$ a.u.\ for the single
layer calculations and of $d=50$ a.u.\ for the double-layer.

Starting from the Kohm-Sham wave-functions and the quasi-particle energies,
the optical-spectra are calculated on the
level of the Bethe-Salpeter equation (BSE):\cite{strinati82,strinati84,
rohlfing2000,palummo2004}
\begin{equation}
(E_{c{\bf k}}-E_{v{\bf k}}) A^S_{vc{\bf k}} +
\Sigma_{{\bf k'}v'c'} \left\langle
vc{\bf k}|K_{eh}|v'c'{\bf k'}\right\rangle A^S_{v'c'{\bf k'}} =
\Omega^S A^S_{vc{\bf k}}.
\label{bethesalpeter}
\end{equation}

Here, the electronic excitations are expressed in the basis of electron-hole 
pairs (i.e., vertical excitations at a given ${\bf k}$-point from a state
in the valence band with quasi-particle energy $E_{v{\bf k}}$ to
a conduction-band state with energy $E_{c{\bf k}}$.
The $A^S_{vc{\bf k}}$ are the expansion coefficients of the excitons
in the electron-hole basis and the $\Omega^S$ are the eigenenergies
(corresponding to the possible excitation energies of the system).
If the interaction kernel $K_{eh}$ is absent, Eq.~(\ref{bethesalpeter})
simply yields $\Omega^S = (E_{c{\bf k}}-E_{v{\bf k}})$, i.e., the
excitations of the system correspond to independent electron-hole pairs.
The interaction kernel $K_{eh}$ describes the screened Coulomb
interaction between electrons and holes, and the exchange
interaction, which includes the so called local fields effect.
$K_{eh}$ ``mixes'' different
single-particle excitations, from valence band states $v,v'$ to conduction
band states $c,c'$, giving rise to modified transition energies
$\Omega^S$ and (possibly) also to discrete excitonic states below the onset
of the continuum.

The optical absorption spectrum is given by the imaginary part of 
the dielectric function, $\varepsilon(\hbar\omega)$, and can be calculated as
\begin{equation}
\varepsilon_2(\hbar\omega)\propto\sum_S\left|\sum_{cv\bf{k}}A^S_{vc\bf{k}}
\frac{\langle c{\bf k}|p_i|v{\bf k}\rangle}{\epsilon_{c\bf{k}}-\epsilon_{v\bf{k}}} 
\right|\delta(\Omega^S-\hbar\omega-\Gamma),
\end{equation}
where $\langle c{\bf k}|{\bf p}|v{\bf k}\rangle$ are the dipole matrix elements
for electronic transitions from valence to conduction states. 
We consider only light absorption with polarization, and thus the direction
of the dipole operator $\bf{p}$, in the plane identified by the $MoS_2$ layer.
The out-of-plane polarization gives negligible contribution to absorption at 
low energies, because
the local fields, which are strongly inhomogeneous in that direction, shift the oscillator
strength to high energies. Excluding phonon-assisted transitions,
the momentum $\bf{k}$ is conserved in the absorption process.
We use an energy broadening $\Gamma=0.05$ eV in all the calculations to mimic
the experimental resolution. The BSE calculations have been performed
with the code {\tt Yambo}.\cite{yambo} Since a much higher $\bf{k}$-point
sampling than for the GW-calculations is needed, we use LDA wave-functions
and energies, corrected by a ``scissor''-operator obtained from the 
GW-calculations. While this approach does not take into account changes
in the valence and conduction band dispersions, we have checked for the
excitonic spectrum at a sampling of $18 \times 18 \times 1$ (single-layer)
that the difference to a BSE-calculation based on GW-energies is negligible.

\begin{figure*}
\begin{center}
\includegraphics[width=\linewidth]{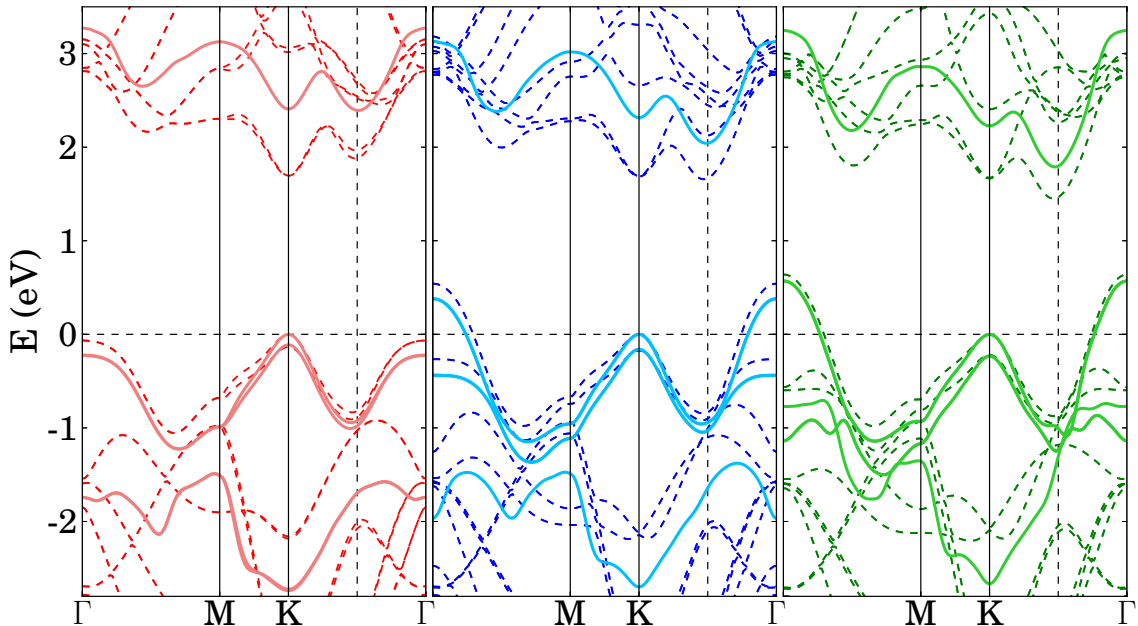}
\caption{(Color online) From left to right: Band structure of  MoS$_2$ single-, double-layer and bulk in the LDA (thin dashed lines)
and in the $G_0W_0$ approximation (thick continuous lines).}
\label{band}
\end{center}
\end{figure*}

\section{Quasi-particle band-structure}
First we study
the electronic structure of single-layer, double-layer and bulk 
MoS$_2$. It is worth to mention that an accurate calculation of the exchange
interaction in both the GW approximation
and the BSE requires the use of semi-core (4s and 4p) orbitals for the 
molybdenum atoms. 
The omission of semi-core states can lead to an erroneous wave-vector
dependence of the GW-correction.
Furthermore, the spin-orbit interaction has to be taken into account
because it removes the degeneracy of the valence band maximum.
Single-layer MoS$_2$ (and, in general, odd number of layers) belongs
to the group of symmetry $D_{3h}$,\cite{alex} which lacks inversion symmetry. 
This absence of symmetry, together with the strong spin-orbit coupling 
of molybdenum $d$ orbitals, splits the valence band edge at $K$.\cite{xiao2012} 
MoS$_2$ with even number of layers, such as the double-layer, 
and the bulk, instead, belong to the symmetry group $D_6h$, which does
have inversion  symmetry. Nevertheless, also in this case a valence band 
splitting exists but it is caused (predominantly) by the interlayer interaction.

Figure \ref{band} shows the band structure, calculated with LDA and GW, for single-, double-layer and bulk MoS$_2$. The only 
case where a direct bandgap is observed is the single-layer, 
either in LDA or in GW, while for double-layer 
and bulk the bandgap is clearly indirect. The origin of this difference is 
the interlayer interaction. A second minimum in the conduction band
lies on the high-symmetry line between $\Gamma$ and $K$ (see vertical
dashed lines). For the single-layer, this minimum is higher in energy
than the one at $K$ but in the double-layer and bulk, the inter-layer
interaction leads to a splitting which shifts the minimum to a lower
energy than the minimum at $K$. 
Moreover, in the double-layer and in the bulk, the valence band edge at $\Gamma$
rises higher than the valence band maximum at 
$\bf{K}$. As a consequence the system moves from
a direct band-gap to an indirect gap semi-conductor.
The GW corrections are generally larger for single-layer MoS$_2$
then for the double-layer and the bulk, due to the smaller dielectric
screening in the single-layer case.

Our band-structures and their interpretation agree very well
with the ones of the recent self-consistent GW calculations,\cite{lambrecht2012}
even though details in the approximations are different: in our study, we
do not use self-consistency in the GW-calculations which seems to add
only a minor energy-shift to the G$_0$W$_0$ calculation. In turn, we 
perform the GW calculations with the full spinor wave-functions while in
Ref.\onlinecite{lambrecht2012} spin-orbit coupling is introduced
after the GW-calculations by rediagonalizing the Hamiltonian matrices.
The values of the direct band gaps at $K$ are summarized in 
Table~\ref{band-table}, combined with the valence band splitting. 
The values are in reasonable agreement with the ones of 
Ref.~\onlinecite{lambrecht2012}. Small differences may be due to the
differences in the approximations and from the different values
of the cell dimensions.

We remark in this context the importance of the lattice constant.
Small variations of the latter can shift the position of the conduction and 
valence band edges. Commonly to other GW calculations, we thus use 
experimental lattice parameters (in-plane lattice constant a=3.15 {\AA}
and c=12.3 {\AA} for bulk\cite{nicklow}) in order to avoid artificial 
strain effects.\cite{rinke2007} 
A similar conclusion is drawn in a recent theoretical study about the 
influence of strain in the bandgap of MoS$_2$, where
strain tends to transform the single-layer MoS$_2$ into an indirect
band-gap semiconductor.\cite{mattheis1973,yun2012}

\begin{table}
\begin{tabular}{lccc}
\hline
\hline
\multicolumn{4}{c}{Material parameters}\\
\hline
                    & 1-layer  & 2-layer  & Bulk   \\
$E_K$ (LDA)         &  1.69    & 1.68     & 1.67   \\ 
$E_K$ (GW)          &  2.41    & 2.32     & 2.23   \\
\hline
$\Delta_v$ (LDA)    &  134.3   & 173.8    & 220.1  \\ 
$\Delta_v$ (GW)     &  112.0   & 160.0    & 230.6  \\
\hline
\hline
\end{tabular}
\caption{Bandgap (in EV) at $K$ point and valence band splitting $\Delta_v$ (in meV), as obtained in LDA and GW.}
\label{band-table}
\end{table}

\section{Optical absorption spectra} 

\begin{figure}
\begin{center}
\includegraphics[width=\linewidth]{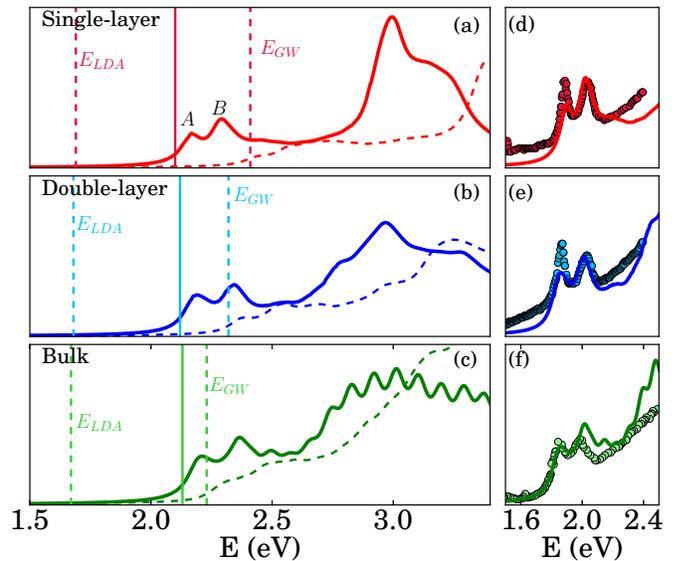}
\caption{(Color online) Panels (a)--(c): Optical spectra for single-layer,
double-layer and bulk MoS$_2$, obtained 
with the BS equation (solid lines) and RPA (dashed lines). 
LDA and GW band gaps at $K$ are marked with vertical dashed lines, the 
absorption threshold is marked with a vertical solid line. Note that the
wiggles above 2.5 eV in panel (c) are due to finite k-point sampling.
Panels (d)--(f):
symbols: experimental absorption spectra\cite{mak2010,splendiani2010}
in comparison with the calculations (solid lines, shifted by about -0.2 eV).
}
\label{bse}
\end{center}
\end{figure}

In the calculation of the optical absorption spectra of MoS$_2$, the 
convergence with respect to  $\bf{k}$-sampling is of crucial 
importance in order to obtain reliable spectra (see Appendix). 
We used a $51\times51\times1$ k-point in the case of the single-layer and 
double-layer and $21\times21\times3$ in the case of bulk. 
Local fields are included in all the calculations.\cite{speedup}
We show in Fig.~\ref{bse}, the calculated optical spectra
for single-, double-layer and bulk MoS$_2$. The results of the BSE are
compared with the optical spectra without the excitonic effects,
calculated in the random-phase approximation (RPA, independent-particle 
picture) using the GW-energies (Panels a-c).
The LDA and GW band-gap energies are marked
with vertical. Finally (Panels d-f) the BSE results are compared with
the experimental data\cite{splendiani2010,mak2010}. In panels d-f
the theoretical results are down-shifted by approximately 0.2 eV in order
to compare with the experiments. This discrepancy is in
the margin error of common GW and Bethe-Salpeter calculations.
Our results give a clear interpretation of the measured absorption
spectra, in the low energy range, and provide reliable predictions
within the given accuracy of $\approx$ 0.2 eV, for the higher energy
range, for which no experimental data are available.

The general features of the three optical spectra are very similar, with a double peak structure (denoted as $A$ and $B$) 
at the energy threshold, accompanied by a plateau, and afterwards an abrupt 
increase of the optical absorption. The RPA spectrum below 2.8 eV resembles 
in all three cases the sum of two Heaviside step functions. The difference
in the two step positions is given by the splitting of the valence band 
maximum at $K$. This step function profile is
the fingerprint of the joint density of states of a 2D-system, with parabolic
band structure.
``Switching on'' excitonic effects, preserves the plateau, but, in addition,
a split excitonic peak below the onset of the continuum transition occurs.
The exciton binding energy is different in the three cases. It is largest
for the single-layer, where the electron-hole interaction is less screened.
The differences in quasi-particle band gap and excitonic binding energy
almost cancel each other\cite{wirtz2006,komsa} and the resulting excitonic 
peak positions are almost the same in the three cases.
The peaks can be directly assigned to the peaks denoted $A$ and $B$
in the experiments.\cite{mak2010,splendiani2010}
At higher energies, the relevant peak placed at 3.0 eV (close to the blue color in the visible 
spectrum) in the single-layer has larger relative intensity with respect to the 
peaks $A$ and $B$. Such intense peak has not yet been reported, as the maximum detection energy in absorption experiments 
is around 2.4 eV.\cite{plechinger2011} 

A further inspection of the excitonic eigenvectors allows to assign to 
each excitonic peak the contributing electron-hole transitions and 
their location at the Brillouin zone. In all the cases, excitons $A$ and $B$ 
come from the energy transition at the $K$ point, even
when the band gap is indirect. This explains the similarity of the experimental
optical absorption. The other relevant peak, at 3.0 eV, comes from the
part of the Brillouin zone between $K$ and $\Gamma$ (marked with a dashed line in Fig.~\ref{band}), where we observe a high density
of states due to the parallel conduction and valence bands.\cite{cardona}

\begin{figure*}
\begin{center}
\includegraphics[width=15 cm]{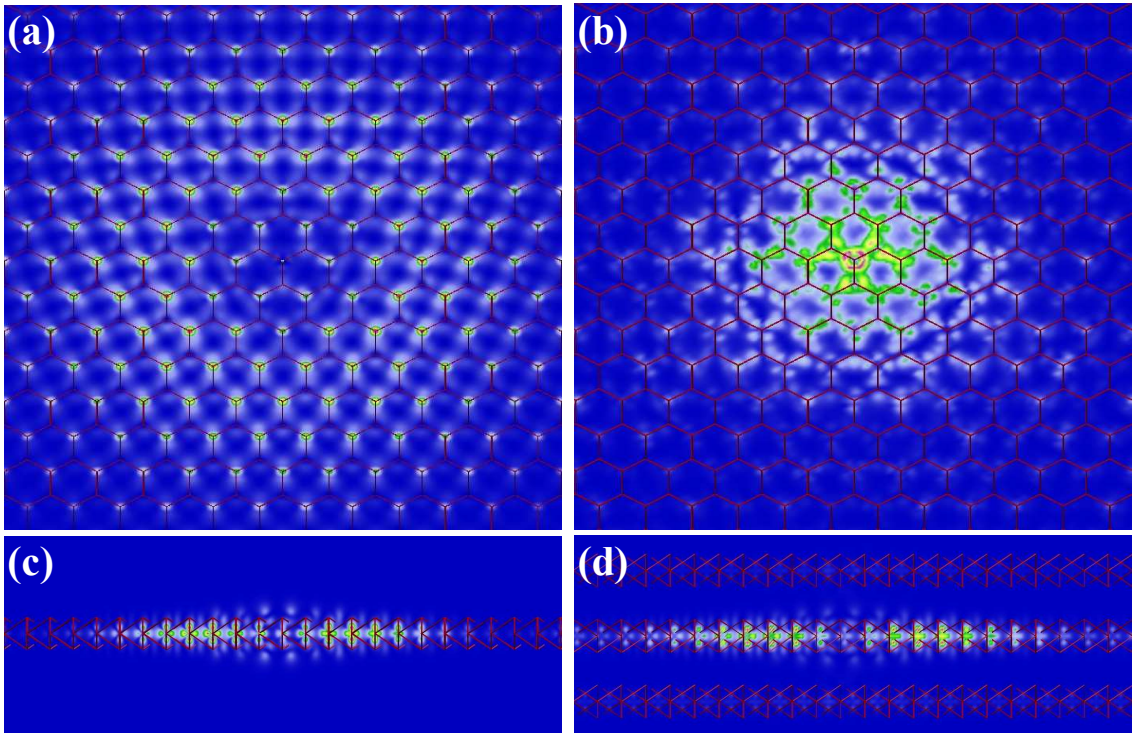}
\caption{(Color online) Exciton wave functions: (a) top view of bound exciton 
$A$ of single-layer, 
(b) top view of bound exciton at 3.0 eV of single-layer,
(c) lateral view of the exciton $A$ of the single-layer, 
(d) lateral view of the exciton $A$ in bulk. 
In all cases, the hole has been placed on the molybdenum atom
in the center of the figure. Images were realized
with the software {\tt XCrySDen}\cite{xcrysden}.}
\label{exciton}
\end{center}
\end{figure*}

To gain further insight of the optical spectra, we represent the excitonic 
wavefunctions related to the main peaks of the Bethe-Salpeter spectra. 
The excitonic wave-function can be written as:

\begin{equation}
|\Psi^S({\bf r}_e,{\bf r}_h)\rangle = \sum_{cv{\bf k}}A^S_{cv{\bf k}}\psi_{c{\bf k}}({\bf r}_h)\psi_{v{\bf k}}({\bf r}_e),
\end{equation}
where $\bf{r}_e$ and $\bf{r}_h$ are the real-space electron and hole coordinate and $\psi$ the LDA wave functions. The 
coefficients $A^S_{cv\bf{k}}$ are obtained by diagonalizing the Hamiltonian of the Bethe-Salpeter equation (with
energy $\Omega^S$). In order to represent the six-coordinate function, we fix the hole position 1.0 {\AA} above a molybdenum atom 
and we project onto the function $|\Psi^S({\bf r}_e, {\bf r}_h=(0,0,1))|^2$ onto
the x-y plane. Figure~\ref{exciton}(a) shows
the exciton wave function of the exciton $A$ of the MoS$_2$ single-layer (for double-layer and bulk the wave function is 
essentially identical). This exciton is largely spread, extended over 65 {\AA}, in concordance with the small
binding energy and with the small effect on the absorption threshold, as reported in the experiments.\cite{mak2010,splendiani2010} This also
explains the large $\bf{k}$-point grid needed to converge the results in the Bethe-Salpeter equation (see Appendix).
The exciton $B$ (not shown here) shows similar trends. 

On the opposite, in Fig.~\ref{exciton}(b) we observe that the brightest exciton, at 3.0 eV, is remarkably 
localized, being confined to less than 30 {\AA}, with a 
trigonal shape. Among the properties of such exciton we point out the potentially high efficiency of recombination. 

Another interesting point (for multi-layer and bulk) is to explore if 
the excitons are confined in one layer or if their wavefunctions
extend over several layers. With this aim we show in Fig.~\ref{exciton}(d) the exciton $A$ of the MoS$_2$ bulk. The wave function spreads largely within the plane but it is undoubtedly constricted to one layer. The density in neighboring
layers is negligible and the wave-function is very similar to the one of
the $A$ exciton in the single layer (Fig.~\ref{exciton}(c)). The large 
interlayer distance, due to the weak van der Waals interaction, prevents the wave function from spreading to other layers, 
analogously to what happens in bulk boron nitride.\cite{arnaud,wirtz2011}

\section{Conclusions} 

We have performed calculations of the quasi-particle band-structure and
of the optical absorption spectra of single, double-layer, and bulk MoS$_2$
including excitonic effects and spin-orbit coupling at the same time.
In agreement with previous calculations and with experimental evidence
from photoluminescence intensities,\cite{mak2010,splendiani2010,sallem2012}
our GW-calculations demonstrate
that only single-layer MoS$_2$ has a direct band gap at $K$. The inter-layer
interaction makes the band gap indirect for multi-layer and bulk Mo$S_2$.
However, this does not particularly influence the optical absorption spectra
in the visible range. These display in all cases a strong excitonic peak 
(composed of electron-hole pairs around $K$) below the onset of continuum
transitions which has the step function shape, corresponds to the 
quasi-2D joint-density of states. 
The layered-structure of the material confines the first exciton
mostly in a single-layer which causes the similarity of the
optical spectrum in all the cases. The position of the split excitonic peak
is remarkably stable with respect to the number of layers (and with
respect to the inter-layer spacing in the calculation for single-layer MoS$_2$).
This has been previously explained as a cancellation effect between the band-gap
correction due to electron-electron interaction and the excitonic binding energy
due to electron-hole interaction\cite{wirtz2006,komsa}.
The splitting of the excitonic peak is directly related to the splitting of the
valence band maximum around $K$ and is entirely due to spin-orbit coupling
for the single-layer and (mostly) due to inter-layer interaction for the
double-layer and bulk.
At higher energy (around 3 eV), the optical spectrum is dominated by
electron-hole pairs from transitions around the center of the line $\Gamma \rightarrow K$ where 
both valence and conduction bands have a minimum and the parallel shape of the bands causes a maximum in the joint density-of-states.
For the single-layer a strongly-bound exciton causes an additional peak in the
spectrum at 3.0 eV. For double-layer and bulk, the shape of the spectrum changes and may allow for a spectroscopic distinction between layer numbers.

\section{Acknowledgments} 

The authors acknowledges financial support from the ANR (French National Research Agency) through project
ANR-09-BLAN-0421-01. Calculations were done at the IDRIS supercomputing center, Orsay (Proj. No.
091827), and at the Tirant Supercomputer of the University of Valencia (group vlc44). 

\appendix
\section{}

\begin{figure}
\begin{center}
\includegraphics[width=7.6 cm]{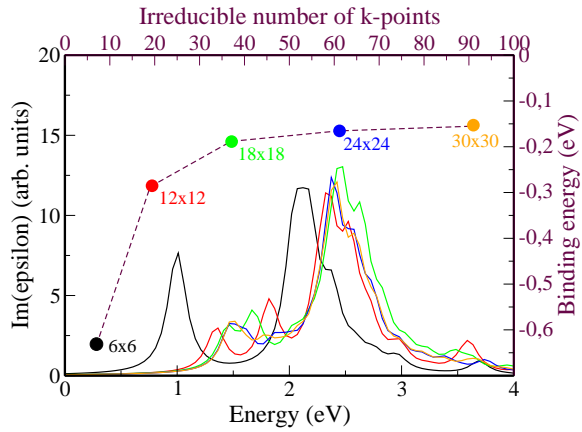}
\caption{(Color online) In the left axis, Bethe-Salpeter spectra for 
several $\bf{k}$-point grid. In the right axis, exciton binding
energy ($E_b$) as a function of the number of irreducible $\bf{k}$-points. Note 
the correspondence in the colors of spectra and $E_b$.}
\label{convergence}
\end{center}
\end{figure}

The convergence of the optical spectra is a mandatory issue to obtain reliable 
spectra, comparable with the experimental data. We have performed calculations 
to check the convergence of the spectra, and in particular of the
position of the first excitonic peak with respect to the
$\bf{k}$-point sampling. A fine sampling can be quite costly 
in terms of computation time and in some Bethe-Salpeter calculations may reach unexpectedly large values. Figure~\ref{convergence}
shows the Bethe-Salpeter spectra of MoS$_2$ single-layer for several $\bf{k}$-grids (for simplicity we 
have omitted the spin-orbit coupling). Additionally, we have represented
the exciton binding energy, $E_b$, as a function of the number of irreducible $\bf{k}$-points,
marking in each case the corresponding grid. We observe the slow convergence of the first excitonic peak, not reached 
before a $18\times18$ sampling. Moreover, for small grids some secondary and artificial peaks appear in the spectrum
and the underlying Heaviside function of the optical absorption is not reached unless dense 
grids are used ($30\times30$). This value is considerably larger than the one used in Ref~\onlinecite{rama} where only a $6 \times 6$ grid was used,
leading to isolated peaks in the absorption spectra where the density of
states predicts a plateau following the excitonic peak. Our test calculations
also explain why in Refs.~\onlinecite{singh}~and~\onlinecite{feng12} split excitonic peaks
 were observed even though spin-orbit coupling was not included.

The exciton binding energy $E_b$ also needs at least an $18 \times 18$ grid to
be converged within 0.1 eV.
It is worth to note that the second excitonic peak (the most intense), located at 
2.5 eV converges much faster, approximately for a $12\times12$ grid. The difference 
behavior of those excitons can be better understood by inspection of their 
wave functions, shown 
in Fig.~\ref{exciton}(a) and (b). In the case of the first peak, the wave function is very spread, more than 18 
unit cells in each direction on the plane. We remind that a $n\times n$ grid in the reciprocal space allows only
to map the exciton wave function in the real space $n\times n$ the unit cell, therefore, inadequately small grids
lead to an artificial confinement of the exciton which increases its binding energy. This arguments also clarifies
why the second exciton converge much faster, its wave function being
confined within a few unit cells. 

\begin{thebibliography}{1}
\bibitem{ppa_note} The exact position of the pole in this model is not important. Moving the pole leads to differences of less than 5 meV.
\bibitem{speedup}
In order to speed up the calculation to obtain smooth spectra, we have also
tested the double-grid method proposed in Ref.~\onlinecite{kammer}. In this 
case a basic grid of $21\times21\times1$ for single-layer and double-layer
yields almost the same results as the $51\times51\times1$ grid.
\end{thebibliography}
\end{document}